# Room-temperature third-order nonlinear anomalous Hall effect in ferromagnetic metal $Fe_3GaTe_2$

Zheng Dai[1], Shuai Zhang[1,2,3,*], Jiajun Li[1], Xiubing Li[1], Congcong Li[1], Fengyi Guo[1], Heng Zhang[1], Ziqi Wang[1], Minhao Zhang[1,2,3], Xuefeng Wang[3,4], Huaiqiang Wang[5], Fengqi Song[1,2,3,*]

[1] National Laboratory of Solid State Microstructures, Collaborative Innovation Center of Advanced Microstructures, Jiangsu Physical Science Research Center, and School of Physics, Nanjing University, Nanjing 210093, China

[2] Institute of Atom Manufacturing, Nanjing University, Suzhou 215163, China

[3] Nanjing Institute of Atomic Scale Manufacturing, Nanjing 211800, China

[4] Jiangsu Provincial Key Laboratory of Advanced Photonic and Electronic Materials, State Key Laboratory of Spintronics Devices and Technologies, School of Electronic Science and Engineering, Nanjing University, Nanjing 210093, China

[5] Center for Quantum Transport and Thermal Energy Science, School of Physics and Technology, Nanjing Normal University, Nanjing 210023, China

*Corresponding authors. S.Z. (szhang@nju.edu.cn), F.S. (songfengqi@nju.edu.cn)

**ABSTRACT**

Berry curvature, as the imaginary component of quantum geometry, plays a crucial role in condensed matter physics. The spatial distribution of Berry curvature can be characterized by its dipole and multipole moments, which can induce the nonlinear anomalous Hall effect (NLAHE). To date, the NLAHE has been demonstrated in various materials, yet reports on room-temperature NLAHE are still limited. In this work, we report the observation of the third-order NLAHE in ferromagnetic metal $Fe_3GaTe_2$. The third-order NLAHE shows hysteretic behavior with the variation of magnetic field, where the coercive field is the same as that of the anomalous Hall effect, and the third-order NLAHE remains observable up to the Curie temperature (~350 K). The scaling analysis suggests that the third-order NLAHE may be attributed to the Berry curvature quadrupole. Our work not only provides an approach to study magnetic materials through nonlinear electric transports, but also opens up possibilities for the future development of room-temperature third-order nonlinear electronic devices.

# Ⅰ. INTRODUCTION

The Hall effect, a fundamental concept in condensed matter physics, has attracted enormous research interest since its discovery. It describes the generation of a voltage perpendicular to the applied current and the magnetic field in a conductive material. In ferromagnetic materials, the anomalous Hall effect (AHE) manifests, enabling the induction of a Hall voltage even in the absence of an external magnetic field. Berry curvature is known to give rise to the intrinsic AHE[1, 2]. A profound comprehension of the quantum nature of the transverse velocity of electrons has further deepened the understanding of the relationship between Berry curvature and the Hall effect in condensed matter physics, and a series of novel Hall effects related to Berry curvature have been discovered, such as the quantum Hall effect[3] and valley Hall effect[4]. These findings not only highlight the central role of Berry curvature in understanding various transport phenomena, but also demonstrate the significance of Berry curvature in condensed matter physics.

In recent years, the second-order nonlinear anomalous Hall effect (NLAHE), as a newly identified phenomenon within the family of Hall effects, has garnered significant attention[5-20]. Distinct from the conventional Hall effect, the second-order NLAHE does not depend on the breaking of time-reversal symmetry but rather on the absence of inversion symmetry[5-7, 12]. The discovery of NLAHE offers a novel perspective for investigating the electronic structure and symmetry properties of quantum materials[12, 13]. The theoretical and experimental studies have demonstrated that the second-order NLAHE can originate from the Berry curvature dipole[5-7, 12, 21-24], the quantum metric dipole[25-27] and extrinsic scatterings[11, 13, 28, 29]. Research on the second-order NLAHE has been progressively deepened, while the explorations of third-order[30-40] and even the higher-order[41] NLAHEs are also advancing rapidly. In particular, the third-order NLAHE has been observed in non-magnetic materials such as $MoTe_2$[30], $TaIrTe_4$[31], $WTe_2$[36] and so on. For the third-order NLAHE in non-magnetic materials, it is generally believed that the physical mechanisms are dominated by the Berry connection polarization tensor[30-32, 38, 42] and quantum metric quadrupole[34, 36, 37]. Recent theoretical studies further indicate that the third-order NLAHE induced by the Berry curvature quadrupole may exist in antiferromagnet and ferromagnet materials[41]. And it has been experimentally confirmed in antiferromagnetic materials, where the third-order NLAHE is associated with the antiferromagnetic transition (e.g., FeSn[35] and $MnBi_2Te_4$[34]). Although significant progress has been made in the research on the third-order NLAHE, reports on room-temperature third-order NLAHE induced by the Berry curvature quadrupole remain rare.

Here, we report the observation of the third-order NLAHE in ferromagnetic metal $Fe_3GaTe_2$. The significant transverse third-harmonic voltage is observed, while the second-harmonic response is negligible. The third-harmonic anomalous Hall voltage satisfies the current dependence of $V_{AH}^{3\omega} \propto (I^{\omega})^3$. It remains observable above room temperature but decreases sharply near the Curie temperature. Through the scaling law analysis, we conclude that the Berry curvature quadrupole is the dominant origin of the third-order NLAHE in ferromagnetic metal $Fe_3GaTe_2$.

## II. EXPERIMENTAL METHODS

*Device Fabrication.* The $Fe_3GaTe_2$ thin flakes were mechanically exfoliated onto $SiO_2$/Si substrate in a glove box with $H_2O$ and $O_2$ levels below 0.1 ppm. The Au electrodes were fabricated by electron beam lithography and electron beam evaporation. The sample thickness was determined by atomic force microscopy (Cypher S). Their exposure to air was limited to <0.5 h. For further protection, hexagonal boron nitride (*h*-BN) thin films were introduced as capping layers on the $Fe_3GaTe_2$ devices.

*Transport Measurements.* Electrical transport measurements were performed in a Quantum Design Physical Property Measurement System with temperatures down to 1.6 K and magnetic field up to 14 T. The voltage difference between different probes was measured using standard lock-in amplifiers (SR 830). The 2-probe DC measurements are performed by using a Keithley 2400. The data shown in the manuscript is collected at a low frequency (47 Hz). A current of 2 mA is employed for all electrical transport measurements in this work, unless pointed out otherwise.

## III. RESULTS AND DISCUSSION

The $Fe_3GaTe_2$ unit cell possesses a hexagonal symmetry with space group $P6_3/mmc$ (no.194), like that of isostructural $Fe_3GeTe_2$, wherein two adjacent quintuple layer substructures with two inequivalent Fe crystallographic sites form a van der Waals (vdW) gap between the tellurium layers, with the structure stacking along the *c*-axis[43, 44], as illustrated in Fig. 1(a). $Fe_3GaTe_2$ has been reported to exhibit a Curie temperature as high as 350~380 K[43, 45], setting a record among known layered vdW intrinsic ferromagnets. Its excellent and robust perpendicular magnetic anisotropy (PMA)[43] and high Curie temperature provide favorable conditions for experimentally exploring the magnetic properties.

We fabricated the $Fe_3GaTe_2$ device via electron beam lithography and electron beam evaporation. The thickness of $Fe_3GaTe_2$ sample is ~33 nm (Fig. S1). The $Fe_3GaTe_2$ device and measurement configuration are shown in Fig. 1(b). By applying an a.c. current $I^\omega$ (frequency $\omega$ = 47 Hz), the longitudinal voltage $V_\parallel$, transverse voltages $V_\perp$ and $V_\perp^{3\omega}$ can be measured simultaneously. Fig. 1(c) shows the temperature dependent longitudinal resistance of $Fe_3GaTe_2$ over a temperature range from 2 K to 360 K. The high-temperature kink corresponds to the ferromagnetic phase transition with $T_C$ ~ 350 K, consistent with previous reports[46-48], confirming that the ferromagnetic properties of $Fe_3GaTe_2$ persist above room temperature. The device exhibits metallic behavior across the majority of the temperature range. However, a deviation toward semiconductor-like behavior emerges around 57 K. This anomaly, also observed in $Fe_3GeTe_2$[49] and $Fe_5GeTe_2$[50], is commonly attributed to weakened electron-phonon scattering in metals[48].

To verify the room-temperature ferromagnetism, we measure the longitudinal and Hall resistance as a function of external magnetic field ($B$) applied in the out-of-plane direction at $T$ = 300 K, as shown in Figs. 1(d) and 1(e). The longitudinal and Hall resistance curves are symmetrized and antisymmetrized by averaging the sum of magnitudes collected in the positive and negative magnetic fields regions to exclude any mixture effect. The anomalous Hall resistance $R_{AH}$ can be obtained from the Hall resistance $R_\perp$ by subtracting the linear background $R_n$, $R_{AH}(B)=R_\perp(B)-R_n(B)$ (Fig. S2). Both the longitudinal resistance and the anomalous Hall resistance display distinct ferromagnetic characteristics. Notably, the Hall rectangular hysteresis loops with near-vertical transitions provide clear evidence of strong PMA[43]. Furthermore, the linear voltage-current relationship displayed in Fig. 1(f) confirms the high-quality ohmic contacts of the $Fe_3GaTe_2$ device.

In addition to the anomalous Hall resistance, we simultaneously characterize the second-harmonic and third-harmonic Hall resistances. The second-harmonic longitudinal and Hall resistances are almost zero (Fig. S4), while the third-harmonic Hall resistance $R_{AH}^{3\omega}$ ($R_{AH}^{3\omega} = V_{AH}^{3\omega}/I^\omega$) exhibits a non-negligible value. Fig. 2(a) shows the variation of the third-harmonic anomalous Hall resistances with magnetic field under different applied currents at 300 K. Similar to the anomalous Hall resistance hysteresis, the third-harmonic anomalous Hall resistance undergoes transitions near the coercive field and exhibits hysteresis behavior. In addition, the third-harmonic anomalous Hall resistance exhibits odd symmetry under positive and negative magnetic fields, which is consistent with that previously reported in antiferromagnetic FeSn[35] and $MnBi_2Te_4$[34]. The magnetic hysteresis phenomenon originates from the ferromagnetic order below the $T_C$. Subsequently, we extract the values of third-harmonic anomalous Hall resistances at different currents under zero magnetic field. The current-dependent third-harmonic anomalous Hall voltages are presented in Fig. 2(b), the black dots are experimental data, and the red solid line denotes the cubic fitting curve. A cubic dependence between the third-harmonic anomalous Hall voltage $V_{\text{AH}}^{3\omega}$ and current $I^{\omega}$ is confirmed, which is consistent with the previously reported phenomena of the third-order NLAHE[35]. While the first-order anomalous Hall resistances (Fig. S3) remain essentially unchanged over the current range. The frequency dependent measurement is shown in Fig. 2(c). The as-obtained third-harmonic anomalous Hall signals are independent of the frequency, demonstrating that the third-order NLAHE is not induced by spurious capacitive coupling effect[7, 12]. Meanwhile, we also rule out the contact junction effect and thermoelectric effect[7, 12, 13] (Figs. S5 and S6).

To characterize the AHE and third-order NLAHE in more detail, we have performed temperature-dependent measurements. No second-harmonic Hall signals are detected throughout the temperature ranges. The magnetic field dependence of the anomalous Hall resistance $R_{\text{AH}}$ and the third-order nonlinear anomalous Hall resistance $R_{\text{AH}}^{3\omega}$ at various temperatures are presented in Figs. 3(a) and 3(b), respectively. The coercive field of the anomalous Hall resistance increases with decreasing temperature, reaching a maximum of 0.45 T at 2 K. With increasing temperature, the anomalous Hall resistance decreases monotonically (Fig. 3(c)). In contrast, the third-order nonlinear anomalous Hall resistance shows a more complex temperature dependence, as shown in Fig. 3(d). The anomalous Hall resistance vanishes around 350 K, consistent with the Curie temperature in Fig. 1(c). Furthermore, we find two characteristic transition temperatures for the third-order nonlinear anomalous Hall resistance: 60 K and 330 K. As can be seen from Fig. 3(d), the third-order nonlinear anomalous Hall resistance $R_{AH}^{3\omega}$ decreases rapidly with increasing temperature from 2 K to 60 K. Above 60 K, it starts to increase and reaches a maximum value at 330 K, then decreases sharply. The temperature-dependent AHE and third-order NLAHE demonstrate that the third-order nonlinear anomalous Hall effect can persist above room temperature and is highly sensitive to the ferromagnetic transitions in Fe$_3$GaTe$_2$.

To further discuss the origin of third-order NLAHE in Fe$_3$GaTe$_2$, we conduct the scaling analysis. We first consider the scaling law relationship of the AHE. As shown in Fig. 4(a), the anomalous Hall resistivity ($\rho_{AH}$) is proportional to the square of the longitudinal resistivity ($\rho_\parallel$), $\rho_{AH} \propto \rho_\parallel^2$. This scaling behavior is consistent with the characteristics of the AHE reported on Fe$_3$GaTe$_2$[45, 51], confirming that the dominant origin of the AHE in Fe$_3$GaTe$_2$ is the Berry curvature. Similar to the scaling analysis of the AHE, the scaling behavior of the third-order NLAHE can be characterized through the temperature dependence of the third-order nonlinear anomalous Hall ($E_{AH}^{3\omega}$) and longitudinal ($E_\parallel$) electric field ratio $E_{AH}^{3\omega}/E_\parallel$ and the longitudinal conductivity $\sigma$. $E_{AH}^{3\omega} = V_{AH}^{3\omega}/W$ and $E_\parallel = V_\parallel/L$, where $L$ and $W$ are the length and width of the channel. As shown in Figs. 4(b) and 4(c), the longitudinal conductivity decreases monotonically at $T > 60$ K, while $E_{AH}^{3\omega}/E_\parallel$ exhibits negligible temperature dependence over a broad range. This is a characteristic feature of the third-order NLAHE induced by the intrinsic Berry curvature quadrupole[35]. To explicitly distinguish the intrinsic and extrinsic contributions (e.g., scattering) to the third-order NLAHE, we adopted the scaling law analysis formula[34, 35] $E_{AH}^{3\omega}/\sigma E_\parallel^3 = \alpha\sigma^2 + \beta$, $\alpha$ denotes the extrinsic contribution and $\beta$ represents the intrinsic Berry curvature quadrupole contribution. We find that when the $\sigma^2$ approaches zero, the Berry curvature quadrupole dominates the NLAHE. As illustrated in Fig. 4(d), our experimental results can be well fitted by the scaling law formula. Notably, the scaling fitting exhibits two distinct regimes. We therefore conducted a detailed analysis to quantify the contributions corresponding to each regime. The fitting of scaling law to the experimental data is shown in Fig. S7, and it yields $\alpha$ = -8.6904×10$^{-29}$ m$^5$Ω$^3$V$^{-2}$ and $\beta$ = 1.56771×10$^{-16}$ m$^3$ΩV$^{-2}$ at 2 K. Considering $\sigma$ = 1.33495×10$^6$ S/m, we find the skew scattering contribution $\alpha\sigma^3$ = -2.06744×10$^{-10}$ m$^2$V$^{-2}$ is comparable with the Berry curvature quadrupole contribution $\beta\sigma$ = 2.09281×10$^{-10}$ m$^2$V$^{-2}$. Nevertheless, as the temperature rises, the Berry curvature quadrupole and skew scattering contribution increase gradually yet the proportion of Berry curvature quadrupole contribution consistently exceeds the scattering contribution. And $\alpha$ = -1.67874×10$^{-30}$ m$^5$Ω$^3$V$^{-2}$, $\beta$ = 3.05974×10$^{-18}$ m$^3$ΩV$^{-2}$ and $\sigma$ =1.19561×10$^6$ S/m at 300 K. Thus, the skew scattering contribution $\alpha\sigma^3$ = -2.86912×10$^{-12}$ m$^2$V$^{-2}$ is much smaller than Berry curvature quadrupole contribution $\beta\sigma$ =3.65824×10$^{-12}$ m$^2$V$^{-2}$ at 300 K. Moreover, the proportion of the Berry curvature quadrupole contribution relative to scattering contribution increases progressively with rising temperature. The above analysis focuses on the situation below 330 K. However, the third-order nonlinear anomalous Hall signal drops sharply near the Curie temperature, a behavior may be attribute to the scattering mechanism[37]. Therefore, we conclude that the intrinsic Berry curvature quadrupole plays the dominant role in third-order NLAHE in Fe$_3$GaTe$_2$.

## Ⅳ. CONCLUSIONS

In conclusion, we observed the third-order NLAHE in $Fe_3GaTe_2$ flakes above room temperature. And a notable transformation is observed near the Curie temperature, indicating that the third-order NLAHE is sensitive to the magnetic transition in ferromagnetic $Fe_3GaTe_2$. Through the scaling law analysis, we suggest that the non-zero Berry curvature quadrupole in $Fe_3GaTe_2$ may dominate the generation of the third-order NLAHE. Our experiment not only offers an approach for probing ferromagnetic phase transitions in materials and understanding the room-temperature third-order NLAHE induced by the Berry curvature quadrupole, but also opens up potential avenues for investigating room-temperature nonlinear devices.

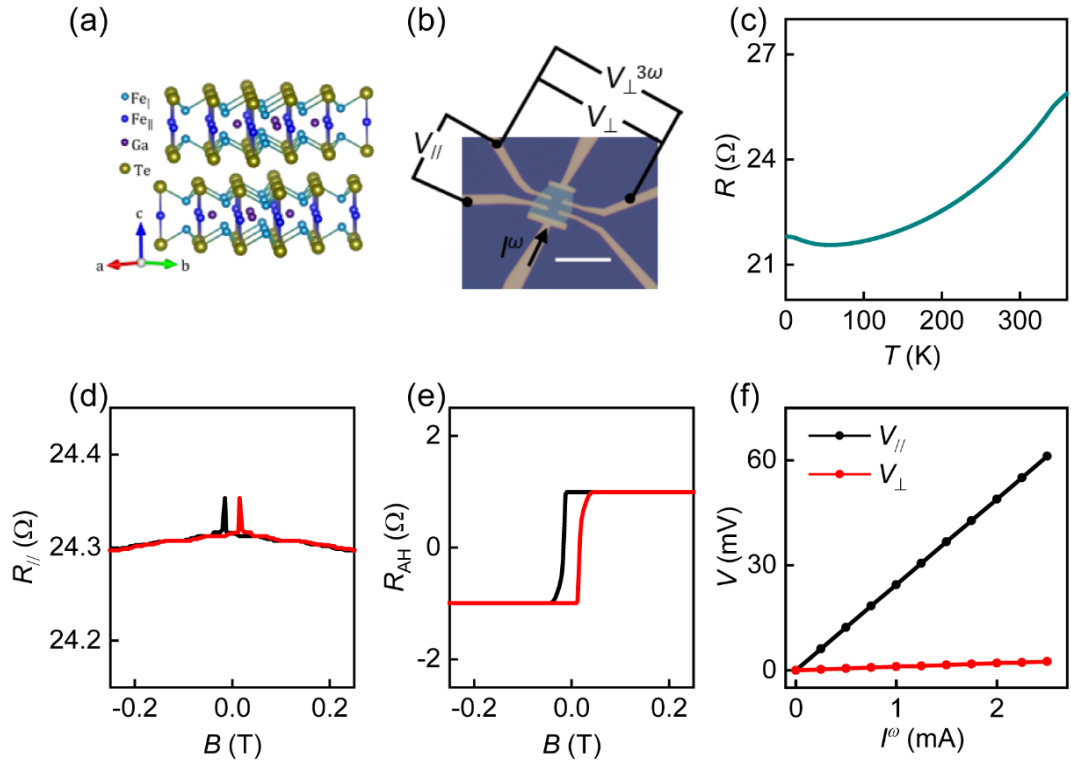

FIG. 1. Basic characterization of the Fe$_3$GaTe$_2$ device. (a) The crystal structure of Fe$_3$GaTe$_2$. (b) The optical image of the Fe$_3$GaTe$_2$ device and the configuration of electrical transport measurement. The white scale bar is 20 μm. (c) The temperature dependent longitudinal resistance. (d)–(e) Magnetic field dependent longitudinal resistance $R_\parallel$ and Hall resistance $R_{AH}$ at 300 K. (f) Current dependent longitudinal voltage $V_\parallel$ and Hall voltage $V_\perp$ at 300 K.

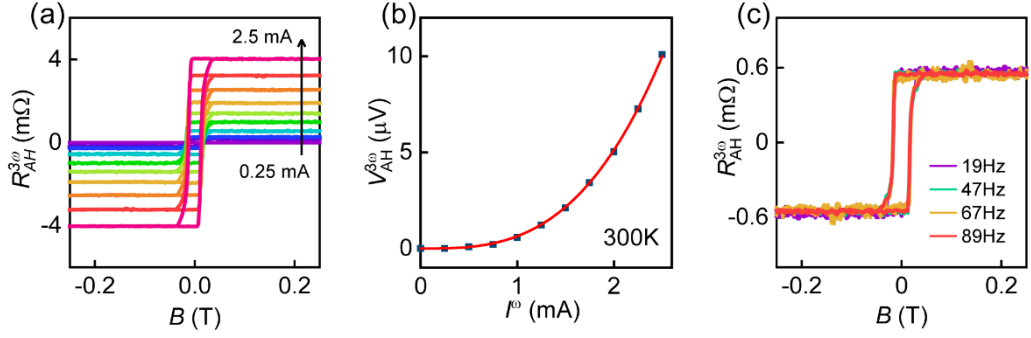

FIG. 2. Room-temperature third-order nonlinear anomalous Hall effect in the $Fe_3GaTe_2$ device. (a) Magnetic field dependent third-order nonlinear anomalous Hall resistance $R_{AH}^{3\omega}$ under different drive currents $I^{\omega}$ at 300 K. (b) Current dependent third-order nonlinear Hall voltages $V_{AH}^{3\omega}$ at $B = 0$ T. The red solid line represents a cubic fit to the experimental data. (c) Magnetic field dependent third-order nonlinear anomalous Hall resistance $R_{AH}^{3\omega}$ under different frequencies with a current of 1 mA.

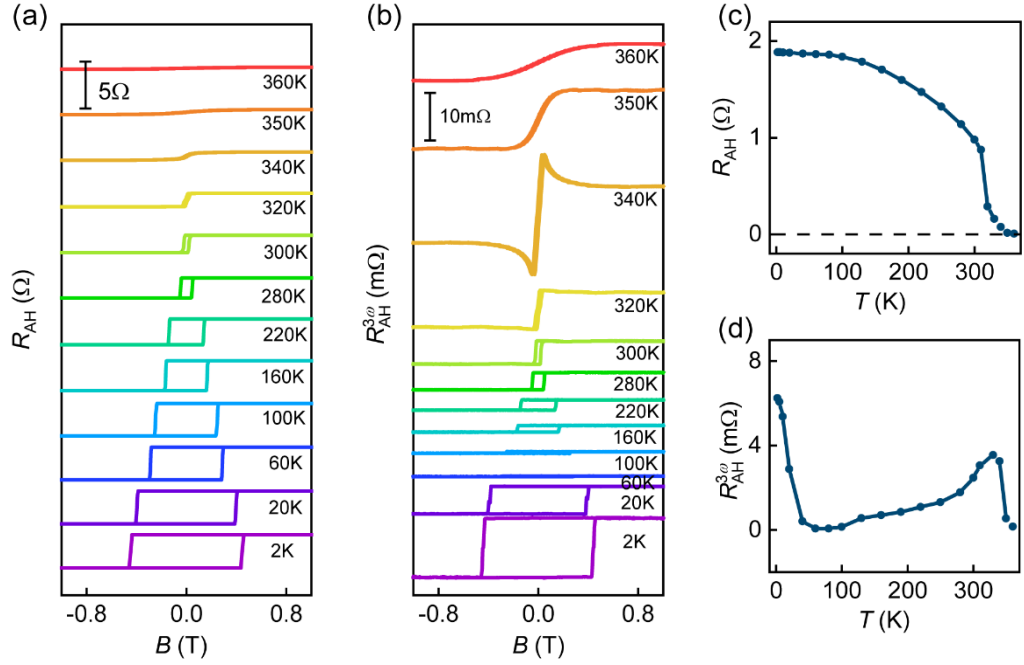

FIG. 3. Temperature dependent anomalous Hall effect and third-order nonlinear anomalous Hall effect in the Fe$_3$GaTe$_2$ device. (a)-(b) Magnetic field dependent anomalous Hall resistance $R_{AH}$ and third-order nonlinear anomalous Hall resistance $R_{AH}^{3\omega}$ at different temperatures. (c) Temperature evolution of anomalous Hall resistance $R_{AH}$. (d) Temperature evolution of third-order nonlinear anomalous Hall resistance $R_{AH}^{3\omega}$ at $B = 0$ T.

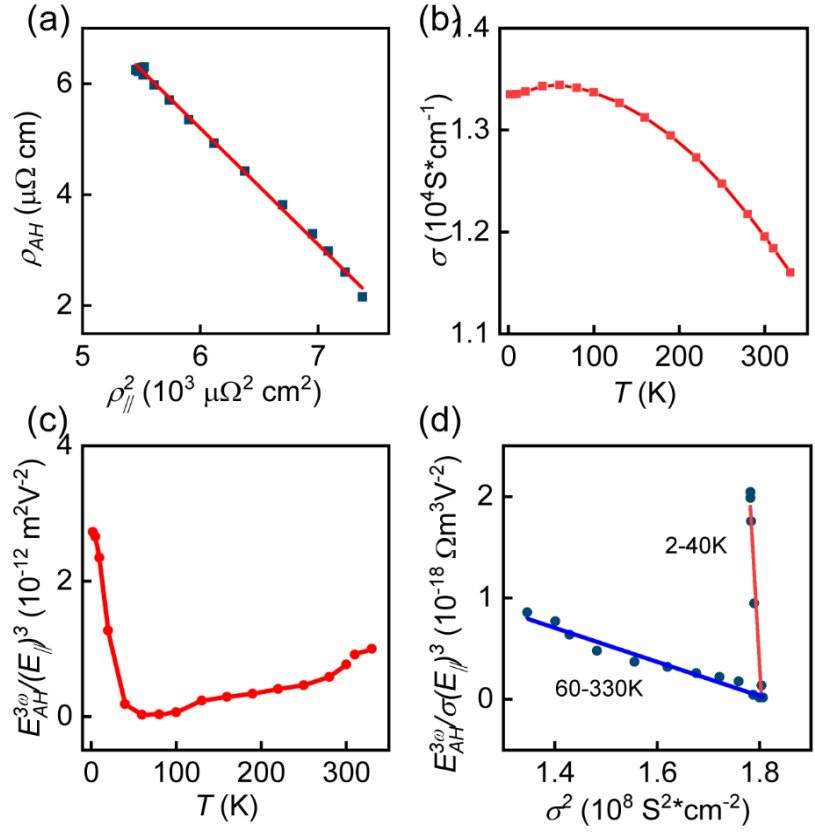

FIG. 4. Scaling law of anomalous Hall effect and third-order nonlinear anomalous Hall effect in the Fe3GaTe2 device. (a) Dependence of $\rho_{AH}$ on the square of $\rho_{\parallel}$. The red solid line represents the linear fit. (b) Temperature dependent longitudinal conductivity. (c) The ratio of the third-order nonlinear anomalous Hall electric field $E_{AH}^{3\omega}$ and the cubed longitudinal electric field $E_{\parallel}$ as a function of temperature. (d) The $E_{AH}^{3\omega}/(\sigma E_{\parallel}^{3})$ and fit to the data (red and blue line) plotted with the square of longitudinal conductivity $\sigma$.